\documentclass[12pt]{interact}

\usepackage{epstopdf}
\usepackage{subcaption}
\usepackage{amsmath,bm,amssymb}
\usepackage{graphicx,xcolor,soul}
\usepackage{tabularx,ragged2e,booktabs}
\usepackage[colorlinks = true, linkcolor = blue, urlcolor  = cyan, citecolor = red, anchorcolor = blue]{hyperref}

\usepackage[numbers,sort&compress,merge]{natbib}
\begin{document}

\title{ Laser-Induced Electron Diffraction:    Alignment Defects and Symmetry Breaking}

\author{
\name{
T.\:T.\:Nguyen-Dang\textsuperscript{1}\thanks{Contact: Thanh-Tung.Nguyen-Dang@chm.ulaval.ca},
	M.\:Peters\textsuperscript{2},
	J.\:Viau-Trudel\textsuperscript{1,3},
	E.\:Couture-Bienvenue\textsuperscript{1},
R.\:Puthumpally-Joseph\textsuperscript{3,4},
E.\:Charron\textsuperscript{3} and
	O.\:Atabek\textsuperscript{3}
}	
\affil{
	\textsuperscript{1}D\'epartement de Chimie, Universit\'e Laval, Qu\'ebec, Canada G1K 7P4\newline
	\textsuperscript{2}Universit\'e de Moncton, Edmundston, NB, Canada E3V 2S8\newline
\textsuperscript{3}Institut des Sciences Mol\'eculaires d'Orsay (ISMO), Univ. Paris-Sud, CNRS, Universit\'e Paris-Saclay, 91405 Orsay cedex, France
\newline
\textsuperscript{4}Laboratoire Interdisciplinaire Carnot de Bourgogne (ICB), UMR 6303 CNRS-Universit\'e Bourgogne Franche Comt\'e, 9 Av. A. Savary, BP 47 870, F-21078 Dijon cedex, France
}
}

\maketitle
\begin{pacscode}
33.80.-b, 34.80.Qb, 34.80.Bm, 42.50.Hz
\end{pacscode}

\begin{abstract}
The fringe pattern that allows geometrical and orbital structure information to be extracted from  LIED spectra of symmetric molecules is shown to reflect a symmetry conservation principle. We show that under a field polarization  which  preserves certain symmetry elements of the molecule, the symmetry character of the initial wave function is conserved during its time-evolution. We present a symmetry analysis of a deviation from a perfect alignment by decomposing   the field into a major, symmetry-determining part, and a minor, symmetry breaking, part. This decomposition leads to a corresponding factorization of the time-evolution operator. The formalism is applied to the analysis of the robustness of LIED readings and inversions with respect to deviations from a perfect perpendicular and parallel alignment of a symmetric ABA triatomic molecule. The results indicate a particularly strong stability of the type of LIED spectra associated with the perpendicular alignment situation.
\end{abstract}

\maketitle
\section{Introduction}
\label{Sec:Intro}
The exploitation of LIED (Laser-Induced Electron Diffraction)  \citep{CPL.259.313, Sc.320.1478, Nat.483.194, NatCom.6.7262} and/or holographic photoelectron spectra  \cite{Bian_ADB_PRL108(2012), Bian_et_al_PRA84(2011)} for dynamic imaging of molecules  \cite{Note_on_dynamic_imaging, Quantum_dyn_imaging_ADB_Ivanov.2011, JPB.49.112001},  as proposed in pioneering works of A. D. Bandrauk and coworkers  \cite{CPL.259.313, Bian_ADB_PRL108(2012)}, requires alignment of the molecule with respect to the linear polarization of the laser field. In one of a series of recent papers dealing with numerical simulations   of LIED data readings and inversions to retrieve  the geometrical and orbital structure  \cite{PRA.83.051403, PRA.94.023421} of a simple, symmetric linear molecule, we found indications of a  rather strong robustness of the LIED imaging procedure  with respect to misalignment  \cite{PRA.85.053417}. This   is a prerequisite for the orbital tomography scheme based on LIED and detailed in Ref.\:\cite{PRA.94.023421} to be valid, as this scheme relied on the inversion of a single LIED spectrum associated with a single alignment situation, as opposed to the tomography scheme based on High-Harmonic Generation (HHG), which assembles HHG spectra for various alignment angles before submitting them to an elaborate inversion procedure \cite{Nat.432.867, Nat.Phys.7.822}.  The robustness of LIED reported in Ref.\:\cite{PRA.85.053417} remains a numerical observation that needs be quantified and understood within  a more complete, formal, quantum mechanical theoretical framework.

The present paper addresses this issue from the viewpoint of symmetry. Symmetry is mostly used in  time-independent quantum mechanics, to classify quantum stationary states of molecular systems, their vibrational modes, and to simplify calculations of matrix elements and integrals \cite{Group_theory_QM_Tinkham.1964, Chem_app_grp_th_Cotton.1990}. It has seldom been evoked in time-dependent quantum mechanics. In discussions and calculations of laser-driven molecular dynamics, correlations have been made between orbital symmetry, or rather, orbital nodal properties, and angular distribution of ionization probability, HHG signals, etc., and these   were rationalized in terms of constructive and destructive interferences of waves emanating, with some coherence, from different sources \cite{Lagmoago_Kamta_ADB:PRA74.033415.2006, Lagmoago_Kamta_ADB:PRA80.041403.2009}. No explicit derivation was made, even in cases the field-molecule interaction still conserves a number of symmetry elements, which together constitute  a subgroup  of the field-free molecule's  point group. This is the type of situations considered in the present paper, as exemplified by the  case of a symmetric linear molecule subjected to a field that is linearly polarized perpendicular to the molecular axis. More precisely, we illustrate the specific case of the HOMO-1 of CO$_2$ for which an inversion procedure for imaging molecular orbital from the photoelectron spectra from the molecule aligned normal to the applied field has been performed in our previous article \cite{MolPhys.in.press}.

We first review some basic facts concerning the commutativity of symmetry operators with the time-evolution operator of the field-dressed molecule, we then demonstrate how the character of the initial wave function (a molecular orbital [MO]) from which the ionized  electron is extracted, is conserved in the exact time-evolution of the quantum system in the presence of the intense laser field.  More precisely, using symmetry group theoretic concepts \cite{Group_theory_QM_Tinkham.1964, Chem_app_grp_th_Cotton.1990}, we show that if the initial state, or a part of it, transforms according to a given irreducible representation of the dressed molecule's symmetry group, i.e. it can be expressed as a definite symmetry-adapted linear combination (SALC) \cite{Chem_app_grp_th_Cotton.1990} of basis functions, then the time-evolved state will continue to transform in the same irreducible representation at all time. The SALC structure of the initial state is conserved, but involves time-dependent functions that result from the propagation of the basis functions (atomic orbitals). It is this time-dependent SALC, TDSALC, structure imparted to the final wave function and its asymptotic form, that gives rise to LIED spectra that carry the signature of the initial orbital and of the molecular geometry.

After reviewing and illustrating these ideas on perfectly aligned molecules, highlighting their relevance to LIED spectra,
we  apply this conservation principle to analyse how the presence of a defect in the molecular alignment affects the symmetry property of the initial MO, and from this, the qualitative traits of the photoelectron spectrum, in particular its part that corresponds to LIED.
Considering the case of a symmetric triatomic molecule of generic formula ABA, We show in particular the stability (or robustness) of the spectra and their inversion with respect to a deviation from a perpendicular alignment of the molecule. Note that we adopt atomic units throughout the paper unless stated otherwise.

\section{Symmetry conservation in a perfect alignment situation}\label{section2}

Consider a symmetric molecule in a linearly polarized laser field. The addition of the laser-molecule interaction usually removes a number of symmetry elements of the field-free molecule, and lowers the symmetry to a subgroup of the isolated molecule's point group. It is with respect to this molecule+field subgroup that the following considerations and conclusions apply.  We will use different notations to designate a symmetry operation, depending on the type of object it acts on. On a physical object, such as a point in real space, or an atom, an electron, we will designate the operation by $\mathcal{S}$. On a vector in three-dimensional position or momentum space, it will be designated as a matrix by $\mathbb{S}$, while on a function, it will be designated as an operator by $\hat S$.

\subsection{Symmetry-related Atomic Orbitals}
Typically, a molecule consists of a number of unique atom types $A, B, C, D,..$ etc. The $n_A$ A-type atoms, $A_1, A_2,..., A_{n_A}$ of position vectors $ \vec R^{A}_1, \vec R^{A}_2, ..., \vec R^{A}_{n_A}$, are related to each other by symmetry operations $ \mathcal{S}$, i.e.
	\begin{equation}
		\exists \  \mathcal{S}_j  \|\ \,\,\,\, \mathcal{S}_j A_1 = A_j,  \qquad
		\mathbb{S}_j \vec R^{A}_1 = \vec R^{A}_j
	\end{equation}
If the nuclear center $A_1$ supports an atomic orbital  (AO)
\begin{equation}
\chi_{l,1}^ A:=\chi_l^A(\vec r-\vec R^{A}_1)=\hat T_{\vec R^{A}_1}\chi_l^A(\vec r)\,,
\end{equation}
Here $\hat T_{\vec \rho}=\exp{(-i\vec \rho. \vec p})$ where, $\vec p=-i \vec \nabla$,  is the translation operator of $\vec r$ by $-\vec \rho$. This member of the AO basis will be repeated (and rotated) by $\hat S_j$, such as to give the same AO centered at nucleus $A_j$:
\begin{equation}\label{AO_rotated}
	\chi_{l,j}^ A:=  \hat S_j \chi_{l,1}^ A
\end{equation}
We recall that the action of the operator $\hat S_j$ on a function $f(\vec r) $ is defined by
$$\hat S_j f(\vec r):= f(\mathbb{S}_j^{-1}\vec r),$$
so that
\begin{equation}\label{AO_rotated_expand}
	\chi_{l,j}^ A(\vec r):=    \chi_{l}^ A(\mathbb{S}_j^{-1}[\vec r-\vec R^{A}_1])=\hat T_{\mathbb{S}_j^{-1}\vec R^{A}_1}\hat S_j \chi_l^A(\vec r).
\end{equation}
Actually, this relation can be generalized for any function $f(\vec r)$, any translation ($\hat T_{\vec \rho}$) of $\vec r$ and any symmetry operator $\hat S$:
\begin{equation}\label{f_translated_rotated}
	\hat S \hat T_{\vec \rho}  f(\vec r)=  \hat T_{\mathbb{S}^{-1}\vec \rho} \hat S f(\vec r)\,.
\end{equation}
By definition, $\hat S$ commutes with the molecular Hamiltonian $\hat H$. Let $\hat U_H(t,t_0)$ be the time evolution operator associated with $\hat H$, i.e.
\begin{equation}\label{U_H_def}
	i\frac{\partial \hat U_H(t,t_0)}{\partial t}= \hat H \hat U_H(t,t_0), \qquad \hat U_H(t_0,t_0)=1\,,
\end{equation}
we have
\begin{equation}\label{U_H_S_commute}
	[\hat U_H(t,t_0), \hat S]=0,
\end{equation}
as can be seen by expanding $\hat U_H(t,t_0)$ in Dyson series, each term of which being a time-ordered power of $\hat H$ which obviously commutes with $\hat S$.
\subsection{SALC and Symmetry-adapted MO}
If $\Gamma$ is a $g$-dimensional irreducible representation of the relevant group (here, the point group of the molecule+field system, a subgroup  of dimension $h$  of that of the field-free molecule), with $(g\times g)$ matrices $\mathbb{G} (\mathcal{S})$ representing the symmetry operators $\hat S$, then a symmetry-adapted linear combination (SALC) of the $\chi_{l,i}^ A$ functions is obtained by the action of the projection operator ($u=1,2,.., g$) \cite{Chem_app_grp_th_Cotton.1990},
\begin{equation}\label{proj_irrep}
	\hat P_{u}^{\Gamma}= \frac{g}{h}\sum_{j} \mathbb{G} _{uu}(\mathcal{S}_j)\hat S_j,
\end{equation}
on $\chi_{l,1}^ A$ :
\begin{eqnarray}
	\chi_{l,u}^{A,\Gamma}(\vec r)  &=& \hat P_{u}^{\Gamma} \chi_{l,1}^ A(\vec r)\nonumber \\
	&=& \frac{g}{h}\sum_{j}  \mathbb{G} _{uu}(\mathcal{S}_j)\hat T_{\mathbb{S}_j^{-1}\vec R^{A}_1}\hat S_j \chi_l^A(\vec r)
\end{eqnarray}
giving a  symmetry-adapted (SA) basis function, actually one of the basis function (the $u$-th one) of the irreducible representation $\Gamma$  generated by the AO $\chi_{l}^ A$ of atom A. If the basis set of atomic orbitals is complete, then these SALC will constitute also a complete basis for one-electron functions of symmetry $\Gamma$. A field-free molecular orbital (MO), of symmetry species $\Gamma$, and considered as  an initial state for a single-ionization event under the field, can be written in the following linear combination of   atomic orbitals (LCAO):
\begin{equation}\label{LC_SA_AO}
	\psi_{u}^{\Gamma}(\vec r ,0) =N\sum_{\Omega=A,B,..}\sum_{l}\sum_{j}  c_{\Omega, l}\mathbb{G} _{uu}(\mathcal{S}_j)\hat T_{\mathbb{S}_j^{-1}\vec R^{\Omega}_1}\hat S_j \chi_l^\Omega(\vec r),
\end{equation}
where $N$ is a global normalization constant.

This expresses the molecular orbital in the form of a   combination of the atomic orbitals  $\hat T_{\mathbb{S}_j^{-1}\vec R^{\Omega}_1}\hat S_j \chi_l^\Omega(\vec r)$. The corresponding  LCAO expansion coefficients are $c_{\Omega, l}\mathbb{G} _{uu}(\mathcal{S}_j)$, and it is the elements  $\mathbb{G} _{uu}(\mathcal{S}_j)$ of the matrices $\mathbb{G}(\mathcal{S}_j)$  representing the   symmetry operations $\hat S_j$ that ensure that  this MO transforms as a basis vector of the irreducible representation $\Gamma$. They give to the MO a definite symmetry-related nodal structure.
\subsection{Conservation of Orbital Symmetry in Time-evolution: TDSALC}
Using Eq. (\ref{f_translated_rotated})     and the commutativity of $\hat U_H(t,0)$ and $\hat S_j$, Eq. (\ref{U_H_S_commute}), it can be shown that $\hat U_H(t,0)\psi_{u}^{\Gamma}$ still transforms in $\Gamma$, and is actually of the same SALC form as the initial state. More precisely, the final exact result is:

\begin{equation}\label{LC_SA_AO_t_evolved}
	\hat U_H(t,0) \psi_{u}^{\Gamma}(\vec r ,0) =N\sum_{\Omega=A,B,..}\sum_{l}\sum_{j}  c_{\Omega, l}\mathbb{G} _{uu}(\mathcal{S}_j)\hat T_{\mathbb{S}_j^{-1}\vec R^{\Omega}_1}\hat S_j \chi_l^\Omega(\vec r, t),
\end{equation}
where, by definition
\begin{equation}\label{t_dep_AO}
	\chi_l^\Omega(\vec r, t)=\hat T_{ \vec R^{\Omega}_1}^{-1}\hat U_H(t,0)  \hat T_{ \vec R^{\Omega}_1} \chi_l^\Omega(\vec r).
\end{equation}
This is the atomic orbital $\chi_l^\Omega$ centered at nucleus $\Omega$ evolved in time, then translated back to the origin of the body-fixed coordinate system.
Note that the definition of this time-dependent orbital   implies the following property, which can  readily be proven using   Eqs.(\ref{f_translated_rotated}) and  (\ref{U_H_S_commute}). Let $\hat S_1, \ \hat S_2$ be two symmetry operators such that $\mathbb{S}_2 \vec R^{\Omega}_1= \mathbb{S}_1 \vec R^{\Omega}_1$, then
\begin{equation}\label{S1Ao_eq_S2Ao}
	\hat S_2   \chi_l^\Omega(\vec r)=\hat S_1   \chi_l^\Omega(\vec r)\Longleftrightarrow \hat S_2   \chi_l^\Omega(\vec r, t)=\hat S_1   \chi_l^\Omega(\vec r, t)
\end{equation}

With Eq. (\ref{LC_SA_AO_t_evolved}), we see that in terms of the  time-dependent AO $\chi_l^\Omega(\vec r, t)$, the time-evolved state clearly has the same structure as  $\psi_{u}^{\Gamma}(\vec r ,0)$ in Eq. (\ref{LC_SA_AO}).

\subsection{The special case of  ABA linear molecules }
\begin{figure}[ht]
	\centering
	\includegraphics{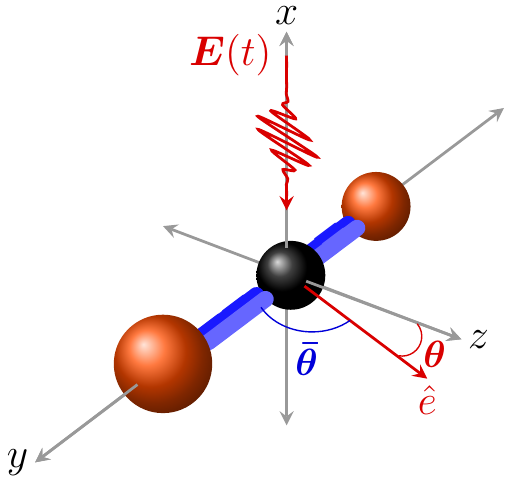}
	\caption{(Color online).   Coordinate system used in the symmetry analysis of 
		the photoelectron spectrum associated with the laser-driven ionization of a CO$_2$ molecule. In the perfect perpendicular alignment situation, the molecule lies along the $y$ axis, while  the polarization of the field lies along the $z$ axis, defining  the two-fold symmetry axis of the $C_{2v}$ subgroup of the molecule+field system. In analyzing the effect of an alignment defect, we will imagine the polarization of the field making an  angle $\theta$ with the $z$ axis, so that this angle measures a deviation from the  perfect perpendicular alignment situation.}
	\label{Fig_CO2_coord_system}
\end{figure}

An ABA linear molecule, such as CO$_2$, is of $D_{\infty h}$ symmetry in field-free condition, and of $C_{2v}$ symmetry in a field linearly polarized perpendicular to the internuclear axis.
To directly make use of the character table of $C_{2v} $ as conventionally given in textbooks, we will call $z$ the direction of the polarization vector, defining the 2-fold symmetry axis, and $y$ the internuclear axis, with the central atom B   found at the origin and the A atoms at $y=-R$ and $y=+R$. This geometry is shown in Figure \ref{Fig_CO2_coord_system} specifically for the CO$_2$ molecule.

Consider the ionization of CO$_2$ from the HOMO-1 orbital (the second highest-occupied molecular orbital of the molecule in its field-free electronic ground-state). This is a $\pi_u$ orbital in $D_{\infty h}$. It becomes a $a_1$ orbital in $C_{2v} $  and, in a minimal basis, is composed of a $2p_z$ atomic orbital centered on each atom. The $a_1$ SALC of the $2p_{z,\rm{O}}$'s, $(\hat 1 + \hat C_2 +\hat \sigma_{xz}+ \hat  \sigma_{yz})\hat T_{-R}2p_{z,\rm{O}}$ is, after evaluating the action of the four symmetry operators on the $2p_{z,\rm{O}}$ orbital, up to a normalization constant:
\begin{eqnarray}\label{A1_SALC_2pzO}
	\rm{SALC}_{a1}(2p_{z,\rm{O}})(\vec r)&=& \hat T_{-R} 2p_{z,\rm{O}}(\vec r) + \hat T_{+R} \hat S 2p_{z,\rm{O}}(\vec r)\\&=&2p_{z,\rm{O}_1}+2p_{z,\rm{O}_2}, \nonumber 
\end{eqnarray}

We have used the facts that $\mathbb{S}^{-1}(-R \vec e_y)=+R \vec e_y$, i.e. $\hat T_{+R}$ is the image of $\hat T_{-R}$ under $\hat S=\hat C_2$ or $\hat \sigma_{xz}$ and, in the second line,
that a $2p_{z,\rm{O}}$ function is of symmetry $a_1$, i.e. it is invariant under all symmetry operations of $C_{2v}$. Eq.(\ref{S1Ao_eq_S2Ao}) then implies that the corresponding  time-evolved function  (defined in Eq. (\ref{2pzO_t}) below) also transforms in $A_1$.

The  HOMO-1 is   given by the combination of this SALC Oxygen orbital with a Carbon $2p_z$ orbital (itself belonging to $a_1$):
\begin{equation}\label{HOMO-1_CO2}
	\psi_{\rm{HOMO-1}}^{\rm{CO2}}\equiv \psi^{a1}(\vec r,0)=a\,\,{\rm{SALC}}_{a1}(2p_{z,\rm{O}})(\vec r) + b\,\,2p_{z,\rm{C}}(\vec r).
\end{equation}
with LCAO coefficients $a$ and $b$. According to Eq. (\ref{LC_SA_AO_t_evolved}), under $U_{H}(t,0)$ this simply becomes
\begin{eqnarray}\label{HOMO-1_CO2_evolved}
	\psi^{a1}(\vec r,t)&=&a\,\,[\hat T_{-R} 2p_{z,\rm{O}}(\vec r,t) +\hat T_{+R}   2p_{z,\rm{O}}(\vec r,t)] + b\,\,2p_{z,\rm{C}}(\vec r,t).
\end{eqnarray}
The time-evolved $2p_z$ functions are
\begin{subequations}
	\begin{eqnarray}
		2p_{z,\rm{O}}(\vec r,t) &=& \hat T_{-R}\hat U_{H}(t,0)\hat T_{+R} \ 2p_{z,\rm{O}}(\vec r) \label{2pzO_t}\\
		2p_{z,\rm{C} }(\vec r,t) &=& \hat U_{H}(t,0) \ 2p_{z,\rm{C} }(\vec r), \label{2pzC_t}
	\end{eqnarray}
\end{subequations}
The momentum distribution derived from  the time-dependent wavefunction of Eq. (\ref{HOMO-1_CO2_evolved}) is readily obtained as the squared modulus of the following amplitude
\begin{eqnarray}\label{HOMO-1_CO2_evolved_k_wfn}
	A(\vec{k})=\langle \vec k|\psi^{a1}( t)\rangle &=&  a\,\,[ \ \langle \vec k|\hat T_{-R} |2p_{z,\rm{O}}( t) \rangle +\langle \vec k|\hat T_{+R}| 2p_{z,\rm{O}}( t)\rangle \ ] + b\,\,\langle \vec k|2p_{z,\rm{C}}( t)\rangle \nonumber \\
	&=&2a\,\,\cos{ ( k_y R)} \langle \vec k| 2p_{z,\rm{O}} ( t)\rangle
	+  b\,\,\langle \vec k|2p_{z,\rm{C}}( t)\rangle.
\end{eqnarray}
Without the $\cos{ ( k_y R)}$   factor  in the   first term, this distribution would be  atomic-like, as it would be the sum of spectra resulting from extraction of an electron from two $2p_{z,\Omega}$ atomic orbitals centered at the origin ($\Omega=\rm{O, C}$).   The $\cos{ ( k_y R)}$ factor modulating the atomic $2p_{z,\rm{O}}$ signal gives rise to a fringe pattern along the $k_y$ direction that is a reflection of the molecular structure, both with respect to its geometry (it permits the reading of $R$), but also with respect to its electronic structure (it reflects the nodeless structure of the $a_1$ HOMO-1 orbital) \cite{MolPhys.in.press}.

\section{Alignment defects and symmetry breaking}\label{section3}
\subsection{General considerations}
We consider now the case where the molecular alignment with respect to the field polarization is not perfect, defining two components of the field: A major component (this could be parallel or perpendicular) $\vec E_{M}(t)=\sqrt{1-\alpha^2} \ \vec \epsilon_{M}  E(t)$, determining the appropriate subgroup  to be used in the symmetry analysis, and a minor one, $ \vec E_{m}(t)=\alpha \vec \epsilon_{m}  E(t)$, ($\alpha < \sqrt{1/2}$), and the (one-electron) Hamiltonian  reads
\begin{subequations}
  \begin{eqnarray}\hat H(t) = \hat H_{M}(t) +\hat V_m(t)\end{eqnarray}
\rm{where,} 
  \begin{eqnarray}\hat H_{M}(t) = \hat H_0 +  \vec r.\vec E_{M}(t)\end{eqnarray}
\rm{and} 
  \begin{eqnarray}\hat V_m(t)=\vec r. \vec E_{m}(t)\,.\end{eqnarray}
\end{subequations}
This  partitioning of the Hamiltonian suggests the following factorization of the propagator $\hat U_H$:
\begin{equation}\label{U_factorized}
	\hat U_H(t,0)\simeq \hat U_M(t,0)\hat U_m^{I}(t,0)
\end{equation}
where $ \hat U_M(t,0)$ is the propagator associated with   $\hat H_M$, i.e. 
	\begin{equation}\label{UM_def}
		i \frac{\partial \hat U_M(t,0)}{\partial t}=\hat H_M(t)\hat U_M(t,0), 
	\end{equation}
	and $\hat U_m^{I}(t,0)$ is that associated with $V_m$ in the interaction picture defined by $\hat H_M$:
	\begin{equation}\label{Um_def}
		i \frac{\partial \hat U_m^{I}(t,0)}{\partial t}=\left [\hat U_M(t,0)^{-1}\hat V_m(t)\hat U_M(t,0)\right ] \hat U_m^{I}(t,0)\,.
	\end{equation}

Consider now   $\hat U_H(t,0)$ acting on an initial wave function $\psi(\vec r, 0)$. This amounts to propagating, under the major, symmetry-group determining field, an effective initial state defined by $\psi(\vec r, 0)$ modified by $\hat U_m^{I}(t,0)$. It is this function
$$\psi_t^{I}(\vec r,0)=\hat U_m^{I}(t,0) \psi(\vec r, 0),$$
that must be decomposed into symmetry components before we can apply the symmetry analysis of the previous section, pertaining to the action of $\hat U_M(t,0)$ within the symmetry subgroup defined by the major electric-field component.
\subsection{Misaligned linear CO$_2$ molecules}
To illustrate the above procedure, consider the ionization out of the HOMO-1 of a CO$_2$ molecule, under a field whose  polarization vector   makes an angle $\theta$ with respect to the direction normal to the molecular axis, as shown in Figure \ref{Fig_CO2_coord_system}.
\subsubsection{Misalignment from a perpendicular configuration}
The major component of the field is then the perpendicular one, i.e. $M\leftrightarrow\perp$, and the minor one is parallel to the molecular axis, $m\leftrightarrow\parallel$.
\begin{equation}
\hat H_\perp =\hat H_0+z \cos{\theta} E(t), \quad \hat V_\parallel(t)=y \sin{\theta} E(t)
\end{equation}

To obtain $\hat V_\parallel^I(t)$, we note that $\hat y^{I}(t)$ and $\hat p_y^{I}(t)$ satisfy equations of motion of the classical form $d\hat y^{I}(t)/dt=\hat p_y^{I}(t),\, \ d\hat p_y^{I}(t)/dt= -\partial_y v_{coul}(\vec r)^{I}(t),$ where $v_{coul}$ is the field-free molecular Coulomb potential. Here the superscript $I$ denotes operators taken in the interaction representation. Given that the momentum distribution should in principle be taken with respect to the asymptotic electron wave function, we may neglect  $v_{coul}$, in the spirit of the widely used Strong-Field Approximation \cite{PhysRevA.74.063404, Reiss_SFA, PhysRevA.78.033412}.
The interaction picture operator $\hat V_{\parallel}^I$ is then
\begin{equation}\label{V_parra_I(t)}
	\hat V_{\parallel}^I(t)=  \beta E(t) \hat y + \beta \ t E(t)\hat p_y
\end{equation}
where $\beta=\sin \theta$, and $\hat U_\parallel^{I}(t,0)$ is of the form  
	\begin{equation}\label{U_parra_I(t)}
		\hat U_\parallel^{I}(t,0) \simeq  e^{-i \beta^2 \eta(t) }e^{ -i \beta F(t) \hat y}e^{ -i \beta G(t) \hat p_y}\end{equation}
	with
	\begin{equation} F(t)=    \int_{0}^{t}dt' E(t'), \quad  G(t)=   \int_{0}^{t}dt' t'E(t'), \quad  \eta(t)= \int_{0}^{t} dt'F(t')G(t')
	\end{equation}

To work out the detailed expression of the transformed  initial state $\hat U_\parallel^{I}(t,0) \psi (\vec r, 0)$, where $\psi $ designates the HOMO-1 of CO$_2$ and is given by Eq. (\ref{HOMO-1_CO2}), care must be exercised to account for the fact that $e^{-i \beta F(t) \hat y}$ does not commute with the translation operators $\hat T_{\pm R}$, nor with the symmetry operations $\hat C_{2}$ and $\hat \sigma_{xz}$ ($\hat S$ in the following). As to the exponential factor containing $\hat p_y$ in $\hat U_\parallel^{I}$, Eq. (\ref{U_parra_I(t)}), it commutes with $\hat T_{\pm R}$, being itself a translation operator (noted $\hat T_{\beta G}$ in the following) of the same type, which therefore  satisfies Eq. (\ref{f_translated_rotated}). Considering all these points, we obtain (for simplicity, we drop the irrelevant global phase factor $e^{-i \beta^2 \eta(t) }$ found on the r.h.s. of Eq. (\ref{U_parra_I(t)}) ),

\begin{eqnarray}\label{U_para_on_psi0}
	\hat U_\parallel^{I}(t,0) \psi (\vec r, 0)&=& a\left \{e^{-i \beta F(t) R}  \hat T_{-R}\left (e^{-i \beta F(t) y} 2p_{z,\rm{O}} (\vec r  -\beta G(t) \vec e_3)  \right ) \right .\\&+& \left . e^{+i \beta F(t) R}\hat T_{+R} \hat S \left ( e^{+i \beta F(t) y} 2p_{z,\rm{O}} (\vec r +\beta G(t) \vec e_3)\right ) \right \} \nonumber \\ &&+  b \left ( e^{-i \beta F(t) y} 2p_{z,\rm{C}}(\vec r-\beta G(t)\vec e_3)\right ).\nonumber
\end{eqnarray}
To apply the symmetry (SALC structure) conservation principle of Eq. (\ref{LC_SA_AO_t_evolved}), we need to re-express this in terms of SALCs.  To this end, the functions $e^{\pm i \beta F(t) y} 2p_{z,\rm{O}} (\vec r  \pm \beta G(t) \vec e_3)$ can be rewritten in terms of two functions, one of symmetry $a_1$, the other  $b_2$ in the molecule+(perpendicular) field  group:
\begin{equation}\label{exp(ialp_F_y)1s_H}
	e^{\pm i \beta F y} 2p_{z,\rm{O}} (\vec r  \pm \beta G \vec e_3)  = f_{a1}^{2p_{z,\rm{O}}}(\vec r) \pm f_{b2}^{2p_{z,\rm{O}}} (\vec r)
\end{equation}
where
\begin{subequations}
	\begin{eqnarray}
		f_{a1}^{2p_{z,\rm{O}}}(\vec r)&=& \cos{(\beta F y)}\pi_{u}^{2p_{z,\rm{O}}}(\vec r, G)+ i\sin{( \beta F y)}\pi_{g}^{2p_{z,\rm{O}}}(\vec r, G), \label{fa11sH}\\
		f_{b2}^{2p_{z,\rm{O}}}(\vec r) &=&   \cos{(\beta F y)} \pi_{g}^{2p_{z,\rm{O}}}(\vec r, G)+i \sin{(  \beta F y)}\pi_{u}^{2p_{z,\rm{O}}}(\vec r, G),  \label{fb21sH}
	\end{eqnarray}
\end{subequations}
and $\pi_{u(g)}^{2p_{z,\rm{O}}}(\vec r, G):= \sqrt{1/2}(\hat T_{+\beta G}\pm \hat T_{-\beta G} ) 2p_{z,\rm{O}}(\vec r)$. (Note that these combinations are here named after their symmetry character in the field-free molecular point group, $D_{\infty h}$).  Substituting these expressions into Eq. (\ref{U_para_on_psi0}), we end up writing $\hat U_\parallel^{I}(t,0) \psi (\vec r, 0)$  in a SALC expansion form, in terms of new atomic orbitals $f_{a1(b2)}^{2p_{z,\rm{O}}}$, $\ e^{-i \beta F(t) y} \ 2p_{y,\rm{C}}$, (the $\chi_l^\Omega$'s of Eq. (\ref{LC_SA_AO})):
\begin{eqnarray}\label{U_para_on_psi0_SALC_form}
	\hat U_\parallel^{I}(t,0) \psi (\vec r, 0)&=& a\left \{ \left ( e^{-i \beta F(t) R}  \hat T_{-R}   +   e^{+i \beta F(t) R}\hat T_{+R} \hat S \right )  f_{a1}^{2p_{z,\rm{O}}}(\vec r) \right .\nonumber \\
	&+& \left . \left ( e^{-i \beta F(t) R}  \hat T_{-R}   -   e^{+i \beta F(t) R}\hat T_{+R} \hat S \right )  f_{b2}^{2p_{z,\rm{O}}}(\vec r) \right \} \nonumber \\ &&+  b \left ( e^{-i \beta F(t) y} 2p_{z,\rm{C}}(\vec r-\beta G(t)\vec e_3)\right ).
\end{eqnarray}
The time-evolution (transformation) of Eq. (\ref{U_para_on_psi0_SALC_form}) under $\hat U_\perp(t,0)$ then gives, according to Eqs.(\ref{LC_SA_AO_t_evolved}) and (\ref{t_dep_AO}), the following final expression:
%

\begin{eqnarray}\label{U_M_m_on_psi0_final}
\hat U_\perp(t,0)\hat U_\parallel^{I}(t,0) \psi (\vec r, 0)&=& a  \ \xi^{{R}}\left [f_{a1}^{2p_{z,\rm{O}}} +f_{b2}^{2p_{z,\rm{O}} } \right ]_t  + b \left [\xi^{y}  2p_{z,\rm{C}}(\vec r-G(t)\vec e_3)\right ]_t\nonumber \\
&=& a  \ \xi^{\rm{R}} \left [\xi^{y}2p_{z,\rm{O}}(\vec r-G(t)\vec e_3)\right ]_t \nonumber \\ &+& b\ \left [ \xi^{y}2p_{z,\rm{C}}(\vec r-G(t)\vec e_3)\right ]_t\,,
\end{eqnarray}
where we have introduced notations
\begin{subequations}
	\begin{eqnarray}
	\xi^{\rm{R}} &=& e^{-i \beta F(t) R}\hat T_{+ R} + e^{-i \beta F(t) R}\hat T_{- R}\,,\\
	\xi^{y} &=& e^{-i \beta F(t) y} \,
	\end{eqnarray}
\end{subequations}
and $\left [ f(\vec r)\right ]_t$ to represent $\hat T_{-R}\hat U_\perp(t,0)\hat T_{+R}f(\vec r)$. Projecting Eq. (\ref{U_M_m_on_psi0_final}) on the plane wave $ |\vec{k}\rangle $, we find
\begin{eqnarray}\label{k_U_MU_m_on_psi0}
	A(\vec{k})&=&\langle \vec k|\hat U_H(t,0) |\psi (  0)\rangle \nonumber\\
	&=& 2a\, \cos{([k_y+\beta F(t) ] R)}\langle \vec k-\beta G \vec e_3|\left [e^{-i \beta F(t) y} 2p_{z,\rm{O}}  \right ]_t\rangle \nonumber \\&& +\ b\,\langle \vec k -\beta G \vec e_3|\left [ e^{-i \beta F(t) y} 2p_{z,\rm{C}} \right ]_t\rangle
\end{eqnarray}
This gives a momentum distribution that is 
isomorphic to the one associated with the perfect alignment case, as described by Eq. (\ref{HOMO-1_CO2_evolved_k_wfn}). It differs from the latter only by (a) a
modification of the atomic amplitudes by the convolution of the Fourier transforms
of $e^{−i\beta F(t)y}$ with that of the atomic orbitals,  displaced, in reciprocal space and along the $k_y$ direction,  by $-\beta G(t)$,  and (b) the shift of the fringe pattern along $k_y$ by $\beta F (t)$.

Note that   when $t=t_f$, the end of a zero-area pulse, $F(t)$ vanishes, and the above modifications reduce to just the displacement of the Fourier transforms of the atomic $2p_z$ orbitals along $k_y$ by a quantity that depends   on $G(t)$, the area of the product of the field with time, which is usually non-zero. The relatively minor character of these modifications we just found, in particular for a zero-area pulse,  is a first
formal indication of the stability (robustness) of the LIED imaging procedure with respect to misalignment of
the molecule.

\subsubsection{Misalignment from a parallel configuration}
In this case the major component of the field is the parallel one,
$$\hat H_\parallel =\hat H_0 +y \sin{\theta} E(t), \quad \hat V_\perp(t)=z \cos{\theta} E(t),$$
and with $\theta \simeq \pi/2$, it is $\cos{\theta}$ which represents the small quantity $\beta$ in the partitioning of the electron-field interaction.

Rigorously, the molecule+field subgroup should be   $C_{\infty v}$   in the case the field is polarized parallel to the internuclear axis ($\theta = \pi/2$ exactly). However, if we restrict our analysis of photoelectron spectra  to 2D, then consideration of the  $C_{2v} $ subgroup is sufficient.
While with the perpendicular   molecule/field arrangement, the two Oxygens are  symmetry-related, this is not the case for the parallel alignment.   Each of the three AO constituting the HOMO-1 of CO$_2$, Eq. (\ref{HOMO-1_CO2}), is actually a $b_2$ SALC orbital   on its own, and the general Eq. (\ref{LC_SA_AO_t_evolved}) gives in this case a momentum-space amplitude of the form
\begin{equation}\label{HOMO-1_COH2_evolved_parallel_k}
A(\vec{k})=	a\,\left(  \langle \vec k| \hat U_{\parallel}(t,0)\hat T_{-R}| 2p_{z,\rm{O}}  \rangle   +    \langle \vec k|    \hat U_{\parallel}(t,0)\hat T_{+R} | 2p_{z,\rm{O}} \rangle \right) + b \,\langle \vec k|\hat U_{\parallel}(t,0) |2p_{z,\rm{C}}\rangle,
\end{equation}
for a perfect alignment. In this case, the information on the geometry of the molecule results from the interference between the wavepackets associated with the separate propagations of the two $2p_{z,\rm{O}}$'s and the $2p_{z,\rm{C}} $ under $ U_{\parallel}(t,0)$. In other words, this results from the waves emanating from the three centers, without carrying explicitly the coherence information of the  antisymmetric ($\pi_u$) combination of the $2p_{z,\rm{O}}$'s, coherence that give rise to the strong geometric factor $\cos{(k_y R)}$ found in Eq. (\ref{HOMO-1_CO2_evolved_k_wfn}) for the perpendicular case.

With a defect from parallel alignment,  one readily identifies $\hat U_M\leftrightarrow\hat U_{\parallel} $. Now, \mbox{$\hat U_{m}^{I}\leftrightarrow\;\hat{U}_{\perp}^{I}$}  will be given by the r.h.s. of Eq. ({\ref{U_parra_I(t)}) but with $z$ replacing $y$ throughout, neglecting again the  Coulomb force term, $-\partial_y v_{coul}$. This $\hat U_{\perp}^{I}$ commutes with $\hat T_{\pm R}$. It brings to the momentum-space amplitude of Eq. (\ref{HOMO-1_COH2_evolved_parallel_k}) just a modification of the atomic orbitals $2p_{z,\rm{O}}$ and $2p_{z,\rm{C}}$ essentially by $e^{ i (zF(t)} e^{ iG(t) \hat p_z } $ corresponding to a lateral momentum and displacement of the center of these functions.
	
This reading does not give much explicit, useful information to assess the stability of the type of photoelectron distribution expected  for parallel alignment. However, a deviation from the perfect parallel alignment does allow  for an analysis of the wave function and the associated momentum distribution in terms of the symmetry considerations made above in which it is the perpendicular component, though minor, which determines the relevant symmetry  group to be used. Thus one expects the qualitative traits of the photoelectron  spectrum associated with a perpendicular alignment to appear even for a situation close to a parallel alignment.  
\section{Illustrations from Numerical Simulations}\label{section4}
The following illustrations make use of results of numerical simulations of the strong-field ionization  process of CO$_2$   using the SAE (Single-active electron) model and wave packet propagation procedure detailed in  preceding papers, in particular in Ref.\:\cite{PRA.85.053417}.

Figure \ref{CO2_HOMO_LIED_vs_align_or_not} shows the LIED spectrum obtained by integrating over $k_x$ the momentum distribution of a photoelectron ionized, under a 3 optical cycle pulse of $\lambda=2100$ nm and $I=1\times 10^{14} \,\rm{W/cm}^2$,  from the HOMO-1 of CO$_2$, at $R=3.0 \,\rm{\AA} $, left panel,  and $R=5.0 \,\rm{\AA} $, right panel. The black solid line in each panel  denotes the perfectly aligned case $\theta=0$. The curves in dotted lines in various colours are obtained with a $\cos^N(\theta)$ distribution of values of $\theta$, with $N$ chosen to give a distribution full width at half-maximum  of $ 5^\circ$ (red dashed line), $ 10^\circ$ (blue dashed-dotted line) and $ 20^\circ$ (green dotted line).

\begin{figure}[ht]
	\centering
	\includegraphics[width=0.75\textwidth]{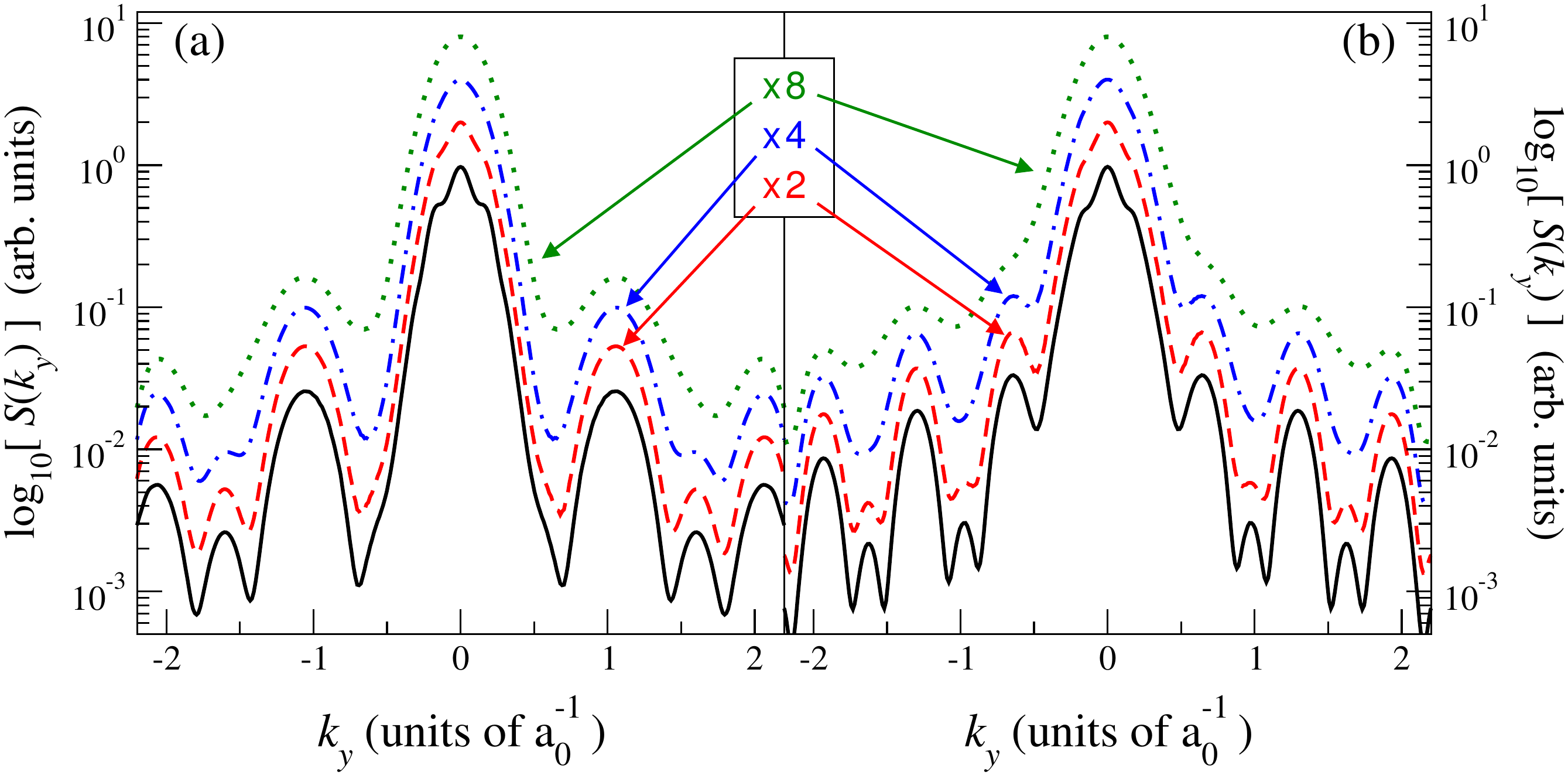}
	\caption{(Color online).  Diffraction patterns $S(k_y)$ obtained by integrating over $k_z$ the high-energy part of the momentum spectrum of a photoelectron ionized, under a 3 optical cycle pulse of  $\lambda=2100$ nm, $I=1\times 10^{14} \,\rm{W/cm}^2$,  from the HOMO-1 of CO$_2$ at $R=3.0\,\rm{\AA} $, panel (a) and  $R=5.0\,\rm{\AA}$, panel (b).  The black solid lines correspond to a perfect perpendicular alignment of the molecule. The coloured dotted lines to a $\cos^N(\theta)$ distribution of values of $\theta$, with $N$ chosen to give a distribution full width at half-maximum of $ 5^\circ$ (red dashed line), $ 10^\circ$ (blue dashed-dotted line) and $ 20^\circ$ (green dotted line). }
	\label{CO2_HOMO_LIED_vs_align_or_not}
\end{figure}

The trace of the photoelectron spectrum along $k_y$ for the perfectly aligned situation, ($\theta=0$),  in both panels (a) and (b) of Figure \ref{CO2_HOMO_LIED_vs_align_or_not} clearly exhibits a fringe pattern with two types of periodicity,    arising from the beating of two terms in the square of  the amplitude $A(\vec{k})$ given in Eq. (\ref{HOMO-1_CO2_evolved_k_wfn}). The product of the  first term  of this equation  with the second one, oscillates with a  period  $2\pi /R$, while the   square of the first term, containing $\cos^2 (k_y R)=[\cos{(2k_y R)}+1]/2$, oscillates with a period of $\pi /R$. The shift of the pure atomic signals (Fourier transforms of the time-evolved $2p_{z, \rm{C(O)}}$) with respect to the geometric $\cos(k_y R)$ factor is responsible for the modifications seen in Figure  \ref{CO2_HOMO_LIED_vs_align_or_not} as a spread of values of $\theta$ is considered. The overall observation is that the lower amplitudes in the spectrum for the perfect alignment case tend to be washed out. Nevertheless, the broad peaks repeating every larger period ($2\pi /R$) are well seen and by themselves allow for a partial reading of the original orbital structure.

The derivations and discussions of the previous section point to the particular stability of the perpendicular alignment type of photoelectron and/or LIED spectrum. It is practically a generic type, i.e. it is expected to be observed even for an alignment situation very close to a parallel one. For the case considered there, of the ionization out of the HOMO-1 of the CO$_2$ molecule, at $R= 3.0 \rm \AA$, Figure \ref{CO2_HOMOm1_photoel_spectra_vs_alignment} shows a sample of photoelectron spectra   obtained for the pulse parameters in Figure \ref{CO2_HOMO_LIED_vs_align_or_not}, as a function of  $\bar{\theta}=\pi/2-\theta$.
\begin{figure}[ht]
	\centering
	\includegraphics[width=0.9\linewidth ]{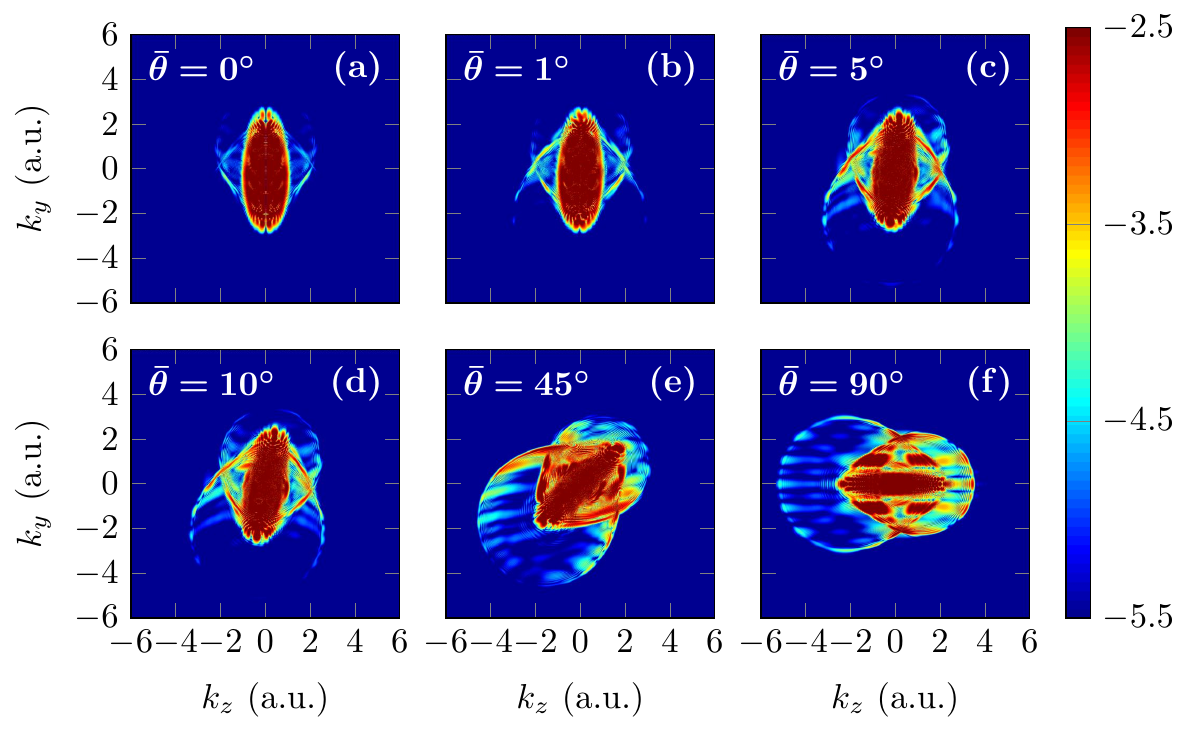}
	\caption{(Color online).  Momentum spectra of a photoelectron ionized from the HOMO-1 of CO$_2$ in a fixed geometry  with $R=3.0\rm{\AA}$ as a function of $\bar{\theta}=\pi/2-\theta$, measuring the deviation from a parrallel alignment of the molecule. $\bar{\theta}=0^ \circ$ in (a), $1^ \circ$ in (b), $5^\circ$ in (c), $10^\circ$ in (d), $45 ^\circ$ in (e) and $90^ \circ$ in (f). The field parameters are the same as in Figure \ref{CO2_HOMO_LIED_vs_align_or_not}}
	\label{CO2_HOMOm1_photoel_spectra_vs_alignment}
\end{figure}

The momentum distribution    of the parallel type (this is at $\bar{\theta}=0^\circ $, panel (a) of Figure \ref{CO2_HOMOm1_photoel_spectra_vs_alignment}) exhibits  a node along $k_z=0$, cutting through the dense  holographic structure found at small $|\vec k|$ and reflecting the nodal structure inherent to atomic $2p_z$ orbitals. This structure is found only  for a narrow range of $\bar{\theta}$, not exceeding $\bar{\theta}=5^{\circ} $. At $\bar{\theta}=5^\circ  $, panel (c), the distribution appears already of the type associated with the perpendicular alignment: the nodal surface that was at $k_z=0$ in the prefect parallel alignment disappears, being replaced by a contour of maximum amplitude, as found at $k_y=0$ in the perfect perpendicular alignment situation, panel (f). The diffraction (R-dependent) fringe pattern at large $|\vec k|$ becomes clearer as $\bar{\theta}$ gradually tends to $\pi/2$. At $\bar{\theta}=  45^\circ $, the distribution is visually the one associated with the perpendicular alignment tilted by $45^{\circ}$ and carries a fringe pattern along $k_y$ that is sufficiently  clear and contrasted to allow for  a useful LIED reading.

These qualitative observations corroborate the predictions of the previous section. Thus, if we consider $\theta$ or $\bar{\theta}$ to be the control parameter, in polar representation of the control plane ($\vec E_\perp, \vec E_\parallel$), then $\bar{\theta}=0^\circ \rightarrow 5^\circ$ defines a region where the photoelectron spectrum is of the type characteristic of a parallel alignment, and the rest of the control plane is associated with the perpendicular alignment. The much larger area of the region of control space associated with the perpendicular alignment denotes the particular stability of this type of spectrum, ensuring a strong robustness to  imaging techniques based on its reading.

\section{Concluding remarks}
The formulations given in Section \ref{section2} and based on symmetry considerations, leading to expressions of the type given in Eq. (\ref{HOMO-1_CO2_evolved}), for the case of the HOMO-1 of CO$_2$, show  that the nodal structure of the time-evolved wave function is inherited from that of the initial orbital by the conservation of its symmetry (of the irreducible representation of this orbital), not from  interferences  of phases  acquired during the time-evolution under the action of the field. The conservation of the nodal structure of the wave function in real coordinate space implies suppressed ionization  in directions parallel to   nodal surfaces of the initial orbital as noted by Lagmago Kamta and Bandrauk \cite{Lagmoago_Kamta_ADB:PRA74.033415.2006, Lagmoago_Kamta_ADB:PRA80.041403.2009}.
In momentum space this conservation principle implies a fringe structure in the photoelectron spectrum  that is presently exploited in LIED readings and inversions. Phase interferences between different electron-wave components acquired during the time-evolution of the system are found in the low-energy part of the photoelectron momentum distribution (the holographic part \cite{Bian_ADB_PRL108(2012), Bian_et_al_PRA84(2011)}) and comes from the evolution of the atomic orbitals composing the MO under the combined effect of the multi-center Coulomb potential and the field.

In deriving the equations describing the conservation of the symmetry character of an orbital, we make use of the time evolution operator $\hat U_H(t,0)$, or $\hat U_M(t,0)$ in Section \ref{section3}, without further expanding on this. We note that  with a decomposition of this as expressed  in Eq. (17) and Eqs.(18a-c) of Ref.\:\cite{PRA.94.023421}, into a sum of propagators associated with free evolution, direct ionization and recollision, the above consideration is applicable to each part of this decomposition, as the corresponding generator commutes with the symmetry operator of the dressed-molecule subgroup.

The dominant correction term  in the symmetry breaking analysis of the effect of a deviation from perfect perpendicular alignment depends on $F(t)$, the time integrated area of the laser pulse, up to time $t$. Thus, to   this first order of the expansion of $\hat V_\parallel^I(t)$, we  see that this correction will be zero at the end of  a physical zero-area pulse. In Section \ref{section4} the derivation  of the result of symmetry-breaking due to misalignment is thus given  in detail to the next order, neglecting, for simplicity, a term that depends on the Coulomb potential. Non trivial symmetry breaking corrections  are obtained that depend  on a new area, non-zero over the full pulse width, that of the product of the field with time.  The same general structure of the momentum distribution shown in  Eq. (\ref{k_U_MU_m_on_psi0})   is expected to be preserved with the correction terms   that depend on the Coulomb force.

The general principles of the symmetry and symmetry-breaking analysis have been easily applied to the case of a linear symmetric triatomic molecule, because in this case the relevant group, $C_{2v} $, has only one-dimensional irreducible representations. We expect that the case of  a polyatomic symmetric molecule,  for which the relevant subgroup admits higher-dimensional irreducible representations, will be much more complex. Future work needs thus be done for the generalization of the present analysis to these cases.  

\section*{Acknowledgment}
R.P.J. and E.C. acknowledge support from  the EU (Project ITN - 2010 - 264951, CORINF).   We also acknowledge the use of the computing cluster GMPCS of the LUMAT federation (FR 2764 CNRS). O.A. acknowledges the organizing committee of the \textit{Andr\'e D. Bandrauk Honorary Symposium on Molecules and Laser Fields} in Orford (QC), Canada, May 2016, for giving him the opportunity of an invited talk partly covering the subject of this article. Financial support from the  Natural Sciences and Engineering Research Council of Canada (NSRRCC) is gratefully acknowledged by T.T.N.D.

\end{document}